# Thermo-dynamic Studies of β-Ga$_2$O$_3$ Nano-membrane Field-effect Transistors on Sapphire Substrate

Hong Zhou†, Kerry Maize†, Jinhyun Noh†, Ali Shakouri†, and Peide D. Ye†*

† *School of Electrical and Computer Engineering and Birck Nanotechnology Center, Purdue University, West Lafayette, IN 47907, USA*

\* Corresponding author

E-mail: yep@purdue.edu    Fax: 765-496-7443

# Abstract

Self-heating effect is a severe issue for high power semiconductor devices, which degrades the electron mobility, saturation velocity, and also affects the device reliability. Applying an ultra-fast and high-resolution thermo-reflectance imaging technique, direct self-heating effect and surface temperature increase phenomenon are observed on novel top-gate β-$Ga_2O_3$ on insulator field-effect transistors. Here, we demonstrate that by utilizing a higher thermal conductivity sapphire substrate rather than $SiO_2$/Si substrate, the temperature rise above room temperature of β-$Ga_2O_3$ on insulator field-effect transistor can be reduced by a factor of 3 and thereby the self-heating effect is significantly reduced. Both thermo-reflectance characterization and simulation verify that the thermal resistance on sapphire substrate is less than 1/3 of that on $SiO_2$/Si substrate. Therefore, maximum drain current density of 535 mA/mm is achieved on sapphire substrate, which is 70% higher than that on $SiO_2$/Si substrate, due to reduced self-heating. Integration of β-$Ga_2O_3$ channel on a higher thermal conductivity substrate opens a new route to address the low thermal conductivity issue of β-$Ga_2O_3$ for power electronics applications.

MAIN TEXT

- **Introduction**

Self-heating effect (SHE) induced temperature increase and non-uniform distribution of dissipated power have emerged as one of the most important concerns in the degradation of transistor's drain current ($I_D$), output power density (P), as well as the gate leakage current, device variability, and reliability[1-5]. This effect would become more severe in high power device with a low thermal conductivity ($\kappa$) substrate such as β-Ga$_2$O$_3$. β-Ga$_2$O$_3$ as a newly emerged semiconductor has an ultra-wide bandgap of 4.8 eV and corresponding high electrical breakdown field ($E_{br}$) of 8 MV/cm, being identified as the next generation wide bandgap semiconductor to replace GaN and SiC[6-9]. In addition, transferrable nano-membrane property will also endow more functionality of the β-Ga$_2$O$_3$ by transferring on different substrate[7]. However, β-Ga$_2$O$_3$ bulk substrate generally suffers from low $\kappa$ of 10~25 W/m·K and thus severe SHE[10,11], which becomes one of major challenges to realize practical applications. An effective approach to mitigate the SHE of β-Ga$_2$O$_3$ field-effect transistors (FETs) is to utilize a higher $\kappa$ substrate rather than the β-Ga$_2$O$_3$ native substrate through a potential wafer bonding technique. In our previous work, we have demonstrated high performance back-gate β-Ga$_2$O$_3$ on insulator (GOOI) FETs by transferring β-Ga$_2$O$_3$ nano-membranes or nano-belts to SiO$_2$/Si substrate with 300 nm SiO$_2$ as the gate voltage blocking layer and Si substrate as the thermal conductor[12,13]. A unique property of β-Ga$_2$O$_3$ is its large lattice constant of 12.23 Å along [100] direction, which allows a facile cleavage into nano-membrane along this [100] direction. Therefore, thin film β-Ga$_2$O$_3$ nano-membrane can be obtained by applying mechanical exfoliation, although β-Ga$_2$O$_3$ is not a van der Waals two-dimensional (2D) material[12].

Nowadays, majority of the GOOI FETs were fabricated on the SiO$_2$/Si substrate like most of 2D devices, such as graphene[14], MoS$_2$[15], and black-phosphorus (BP)[16], however, SiO$_2$ has a low room temperature $\kappa$ of 1.5 W/ m·K. To reduce the SHE of GOOI FET, a higher $\kappa$ substrate is needed. Although SiC has a high $\kappa$ of 300 W/ m·K, it is not an ideal substrate for β-Ga$_2$O$_3$ because of its lower bandgap of 3.3 eV and high cost[17]. Diamond is a promising substrate for β-Ga$_2$O$_3$ with high thermal conductivity, once its high cost and large size unavailable issues are solved[6]. Among those commercial, large size, and wide bandgap substrates, sapphire stands out to be a promising candidate for GOOI FETs, because of its low cost, wide bandgap of 8.8 eV, and medium $\kappa$ of 40 W/ m·K[18], which is around 2-4 times of bulk β-Ga$_2$O$_3$ substrate and around 25 times of SiO$_2$. In this letter, we have used sapphire substrate as a thermal conductor for GOOI FETs to enhance the thermal dissipation and boost the electrical device performance significantly. An ultra-fast, high-resolution thermo-reflectance (TR) imaging technique is applied to the top-gate GOOI FET to examine the reduced SHE and local surface temperature increase (ΔT) above room temperature on sapphire substrate compared to SiO$_2$/Si substrate. After replacing SiO$_2$/Si substrate with sapphire substrate, the top-gate GOOI FET has 70% higher maximum I$_D$ (I$_{DMAX}$), ~ twice lower device surface ΔT, and twice lower thermal resistance (R$_T$).

## ▪ Results and Discussions

**β-Ga$_2$O$_3$ materials and GOOI FETs.** The experimental device cross-section schematic of top-gate GOOI FETs on SiO$_2$/Si and sapphire substrates are shown in Fig. 1(a) and 1(b), consisting of a Ni/Au (30/50 nm) top gate, 15 nm amorphous aluminum oxide (Al$_2$O$_3$) gate dielectric and 70-100 nm Sn doped (100) β-Ga$_2$O$_3$ channel with surface roughness of 0.3 nm[19], and Ti/Al/Au (15/60/50 nm) source/drain contacts. A 270 nm thick of $SiO_2$ is used as drain voltage blocking layer for top-gate GOOI FETs and $p^{++}$ Si is used as a thermal conductor for GOOI FET. The thickness of c-plane sapphire substrate is 650 μm, which is served as drain voltage blocking layer and thermal conductor due to its wider bandgap and higher $\kappa$ compared to (100) β-$Ga_2O_3$ channel. The $\kappa$ of each material is also marked in Fig. 1(a) and 1(b) for comparison. Fig. 1(c) is the false-colored scanning electron microscopy (SEM) image of a fabricated GOOI FET. The device has a gate length ($L_G$) of 1 μm and source to drain spacing ($L_{SD}$) of 6 μm to allow high drain bias to generate enough heat for our TR study. This TR imaging technique offers large-area imaging capability with higher spatial resolution (submicron) compared to other optical thermal characterization approaches, such as infrared (IR) thermography and micro-Raman. TR system also enables a high resolution transient temporal imaging to study the heating/cooling phase at submicrosecond time-scales[20,21].

**I-V Electrical Characteristics.** Fig. 2(a) and 2(b) show the well-behaved direct-current (DC) output characteristics ($I_D$-$V_{DS}$) of two top-gate GOOI FETs with $L_{SD}$ of 6/6.5 μm, $L_G$ of 1 μm, and channel thickness of 73/75 nm on $SiO_2$/Si and sapphire substrates, respectively. The typical range of physical width of nano-membrane devices is 0.5~1.5 μm, determined by SEM as shown in Fig. 1(c). The measurements start from applying the top-gate bias ($V_{GS}$) to 8 or 6 V and then stepping to the device pinch-off voltage of −28 V with −2 V as a step, while the drain bias ($V_{DS}$) is swept from 0 to 27 or 30 V for the $SiO_2$/Si and sapphire substrate devices, respectively. $I_{DMAX}$ of 535 mA/mm and 325 mA/mm for sapphire and $SiO_2$/Si substrates are obtained, the $I_{DMAX}$ on sapphire is the highest among all DC values of all top-gate β-$Ga_2O_3$ MOSFETs[6-9, 22, 23]. $I_{DMAX}$ of GOOI FET on sapphire substrate is around 1.7 times of that on SiO$_2$/Si substrate, originating from better transport properties at a lower device temperature. Fig. 2(c) depicts the I$_D$-V$_{GS}$ and g$_m$-V$_{GS}$ of GOOI FETs on two substrates at V$_{DS}$=25 V. Similar on/off ratio of 10$^9$ and low subthreshold slope (SS) of 65 mV/dec are achieved on both devices due to the wide bandgap and high-quality ALD Al$_2$O$_3$/β-Ga$_2$O$_3$ interface, yielding a low interface trap density (D$_{it}$) of 2.6×10$^{11}$ eV$^{-1}$·cm$^{-2}$ by the equation SS=60×(1+qD$_{it}$/C$_{ox}$) mV/dec at room temperature, where C$_{ox}$ is the oxide capacitance. The slightly high off-state current is from the small device width and limited resolution (10$^{-13}$~10$^{-14}$ A) of I-V measurement equipment. One interesting phenomenon is that the peak g$_m$ of sapphire substrate is 21 mS/mm, which is 60% higher than the g$_m$ of SiO$_2$/Si substrate. The extrinsic electron field-effect mobility (μ$_{FE}$) of GOOI FET on sapphire and SiO$_2$/Si are roughly extracted to be 30.2 cm$^2$/Vs and 21.7 cm$^2$/Vs, respectively, benefiting from the less SHE on sapphire[24]. More than 10 devices on each substrate were characterized and their mobilities were extracted with similar values, confirming the higher mobility on sapphire substrate and the high uniformity of those fabricated devices.

**Steady State TR Imaging.** For power electronics applications with large I$_D$ and V$_{DS}$, the SHE must be taken into consideration. The power consumption (P) of active device can be simply estimated as a product of I$_D$ and V$_{DS}$, then the heat generated in an active device is proportional to the device power consumption. In a GOOI FET, the device serves as the heat source and substrate acts as the heat sink to dissipate the device heat. Heat can also be dissipated through top surface by air convection and source-gate-drain metal pads.[4] However, the heat transfer to the air is very low, considering the low heat transfer coefficient h~10 W/m$^2$/K. A heat-flux (F) of 4.6 × 10$^{-10}$ W can be roughly estimated by equation F=hAΔT, where A is the nano-membrane area of 1 μm × 11 μm, ΔT (43 K) is the device temperature rise above room temperature while being biased for TR measurement. This heat flux is negligible compared to the device power consumption of P= $I_D \times V_{DS}$ =30 V×540 mA/mm ×1 μm = $1.62 \times 10^{-2}$ W, which is mainly dissipated through the substrate. During the TR characterization, the gate electrode metal Au is illuminated through an LED and the reflected signal is calibrated with Au TR coefficient to translate into the increased temperature ΔT above room temperature. The gate Au electrode rather than source/drain contacts is selected, since source/drain contact regions have $Al_2O_3$ on top and the Ti/Al/Au are alloyed. Au directly above the channel region defines the channel temperature and the Au temperature is marked as the channel temperature.

Fig. 3(a) and 3(b) show merged optical and TR thermal image views of GOOI FETs on $SiO_2$/Si and sapphire substrates at steady state and different P conditions. The β-$Ga_2O_3$ nano-membrane has a thickness of 100 nm and 80 nm on $SiO_2$/Si and sapphire substrates, respectively. The 20 nm nano-membrane difference only contributes to a negligible thermal resistance of $2 \times 10^{-6}$ $mm^2 \cdot K/W$, which is less than 0.1% of the total resistance. In this measurement, $V_{GS}$ is biased at 0 V while $V_{DS}$ is swept from 0 V to 27 V or 30 V for GOOI FETs on $SiO_2$/Si and sapphire, respectively. During the TR measurement, the $V_{DS}$ modulation signal has a pulse width of 1 ms and 10% duty cycle. The 1 ms pulse is long enough to allow the FET to reach hot steady state during the high portion of signal. The probe optical pulse width was 100 μs, synchronized just prior to the falling edge of the device excitation pulse to capture the TR image at the end of the $V_{DS}$ pulse. The $I_D$ is recorded at the same time to calculate the normalized power density P=$V_{DS} \times I_D$/A to avoid different heat dissipation area from different size of the β-$Ga_2O_3$ nano-membrane. The area of β-$Ga_2O_3$ nano-membranes are $1.4 \times 10^{-5}$ and $1.1 \times 10^{-5}$ mm$^2$ on SiO$_2$/Si and sapphire, respectively. As higher V$_{DS}$ is applied, device is heated up simultaneously and the corresponding ΔT is increased. At a P = 717 W/mm$^2$ on SiO$_2$/Si substrate the ΔT is measured to be 106 K, while in contrast the ΔT is just 43 K at a higher P = 917 W/mm$^2$ on sapphire substrate. Fig. 3(c) depicts the measured and simulated ΔT versus P for GOOI FETs on two substrates. For both the measurement and simulation results, the clear observation is that at the same P, the GOOI FET on sapphire substrate has more than twice lower ΔT compared to that on SiO$_2$/Si. The R$_T$ of sapphire and SiO$_2$/Si substrates are calculated to be $4.62 \times 10^{-2}$ and $1.47 \times 10^{-1}$ mm$^2$K/W, respectively, through R$_T$ = ΔT/P. The reduced R$_T$ of GOOI FET on sapphire demonstrates that a higher $\kappa$ substrate can be more effectively to dissipate the heat on the devices. With less heat on the device, the temperature is lower so that the μ is higher to achieve a potential higher I$_{DMAX}$ for a better device performance. This substrate thermal engineering approach can also be applied to other 2D FET research if heat dissipation is an issue.

**Transient TR Imaging.** While there are well-established methods to access the DC temperature of devices, there is as yet little direct experimental information available on the device dynamics. Such time-dependent information is particularly relevant for this new device system β-Ga$_2$O$_3$. Understanding of heat dissipation and thermal dynamics of the GOOI FETs are furthermore important because pulse width modulation is commonly used to suppress self-heating and improve efficiency in modern power applications. Dynamic measurements help to reveal the surface temperature redistribution and allow device thermal time constant for both heating (τ$_H$) and cooling (τ$_C$) to be measured experimentally. Measured thermal time constants can be used to approximate device transient thermal resistance based on a simplified 1-dimensional (1D) thermal equivalent circuit model[4]. To precisely determine the thermal time constant, the probe optical pulse width is reduced from 100 μs to 50 ns, and the $V_{DS}$ pulse width is reduced to 1 μs. The first 1 μs heat up phase corresponds to the state when $V_{DS}$ pulse is on and the rest 1 μs cool down phase represents the state when $V_{DS}$ pulse is set to be 0 V. Because the thermal time constants are independent of the amplitude of heating, $V_{DS}$ was chosen to produce similar transient heating amplitude for the GOOI FETs on the two different substrates to avoid different temperature induced variations. Fig. 4(a) and 4(b) describe the three-dimensional (3D) plot ΔT of the GOOI FETs on $SiO_2$/Si and sapphire substrates during the heating and cooling dynamics. Fig. 4(c) is the heating and cooling phase comparison between two substrates. It takes GOOI FETs $\tau_H$ of 350 and 350 ns to reach the steady state during the heat up phase and $\tau_C$ of 250 and 300 ns to cool down on $SiO_2$/Si and sapphire substrates. The cooling phase is of particular importance to investigate the heat dissipation through substrate. Here, we take GOOI FET on sapphire substrate as an example. The time constant of heat dissipation or cooling phase $\tau_C$ through a material is $t^2/(\kappa/\rho C_V)$[4], where, t is heat dissipation depth, ρ is mass density, and $C_V$ is the specific heat. Those parameters are listed in supporting information (SI). For sapphire substrate, the t is calculated to be 2.03 μm at $\tau_C$ =300 ns, which yields an overall transient $R_T$ of $5.0\times10^{-2}$ $mm^2$K/W by $R_T=t/\kappa$[25]. The extracted $R_T$ from transient is in good agreement with the steady state $R_T$ of $4.62\times10^{-2}$ $mm^2$K/W. As for $SiO_2$/Si substrate, this simplification overestimates transient $R_T$ since part of the heat can diffuse up to the gate metal and then get dissipated through lateral diffusion, due to the extremely low κ of $SiO_2$.

**Simulation Verification.** To rationalize our experimental results and evaluate the improvement in the thermal management of GOOI FETs using sapphire substrate to replace $SiO_2$/Si substrate, we have also carried out the modeling work to investigate the SHE on the devices with finite-element method (FEM). In the model, we have defined the heat source as the 80/100 nm thick β-$Ga_2O_3$ nano-membranes. The substrate thickness is set to be 10 μm and the bottom of substrate is set to be T=300 K. Thermal boundary resistance ($R_{TB}$) for β-$Ga_2O_3$ systems are not well known, consequently, we use $R_{TB}$ as a fitting parameter in our model. For an initial value of $R_{TB}$ we use $1.67 \times 10^{-2}$ $mm^2K/W$ between β-$Ga_2O_3$ and two substrates. The choice of GaN $R_{TB}$ for initial value is arbitrary, based only on the prevalence of GaN in power applications[26]. Fig. 5(a) shows the simulated T distribution of GOOI FET on $SiO_2$/Si and sapphire substrates at P of 717 and 917 $W/mm^2$, corresponding to the respective power applied for the experimental TR images. For GOOI FET on $SiO_2$/Si substrate, β-$Ga_2O_3$ nano-membrane has a ΔT=106 K at gate region, in great contrast to the ΔT of 41 K on sapphire substrate, showing the significantly reduced device temperature after implementing a higher κ substrate. Fig. 5(b) shows the transient TR measured ΔT distribution and corresponding FEM simulation of the temperature distribution along gate width across the channel direction on sapphire substrate. At all time regimes, both the simulated and measured temperature profiles coincide very well and indicate that the peak temperature is located at the center of the gate region with β-$Ga_2O_3$ nano-membrane underneath. This profile possesses a sharp decrease of the temperature at the front and back edges of the β-$Ga_2O_3$ nano-membrane, which is normalized to the current conduction direction. This in return verifies our 1D approximation is reasonable that most of the heat is directly dissipated through the substrate rather than being laterally dissipated through the gate metal. On the contrary for GOOI FET on $SiO_2$/Si substrate, there is a significant amount of heat diffused to gate metal and then laterally dissipated through the gate metal region without the nano-membrane underneath, as indicated in Fig. 5(a). Therefore, the 1D simplified model has its limitation to apply on the SiO$_2$/Si substrate or other low $\kappa$ substrates. Finally, good agreements are achieved both on the steady state and transient heating and cooling phases, validating our experiments and the TR characterization technique, and further confirming the reduced SHE using sapphire as substrate.

## ▪ Conclusion

In summary, we have demonstrated that sapphire substrate can provide much better thermal dissipation and less SHE than SiO$_2$/Si substrate for high power device GOOI FETs. Both the simulation and TR imaging reveal that the GOOI FET on sapphire substrate has twice lower $\Delta T$ compared to that on SiO$_2$/Si substrate for identical device power density. The $R_T$ is extracted and calculated to be 4.6×10$^{-2}$ and 1.47×10$^{-1}$ mm$^2$K/W for sapphire and SiO$_2$/Si substrates, respectively. Benefiting from the enhanced heat dissipation, better electronic device performance of GOOI FET on sapphire substrate is achieved.

## ▪ Experimental Methods

**Device fabrication.** Sn doped (-201) β-Ga$_2$O$_3$ bulk substrate was purchased from Tamura Corporation, which has a carrier concentration of 2.8×10$^{18}$ cm$^{-3}$ determined by Capacitance-Voltage (C-V) measurement. The nano-membranes were mechanically exfoliated from the bulk substrate's edge cleavage through scotch tape method. The exfoliation process was repeated to get thin nano-membranes. After the exfoliation, nano-membranes were transferred to the SiO$_2$/Si and sapphire substrates, the same method to get 2D flakes. Prior to the transfer, the SiO$_2$/p$^{++}$ Si and sapphire substrates were cleaned in acetone for 24 hours and the β-Ga$_2$O$_3$ nano-membrane transfer time was within 1 minute. ZEP 520A was used as the electron-beam lithography (EBL) resist in our experiment. Source and drain regions were defined by the VB6 EBL system followed by resist development, Ti/Al/Au metallization and lift-off processes. 30 s of Ar plasma bombardment with radio frequency (RF) power of 100 W was adopted to the source and drain regions before metallization to improve the contact formation. The device without Ar bombardment shows more severe Schottky like behaviors. 15 nm of $Al_2O_3$ was then deposited by an ASM F-120 atomic-layer-deposition (ALD) system at 250 °C with tri-methyl-aluminum (TMA) and $H_2O$ as precursors. Finally, Ni/Au were deposited as the gate electrode, followed by lift-off process.

**Device Characterization.** The thickness and surface roughness of the (100) $\beta$-$Ga_2O_3$ channel was measured by a Veeco Dimension 3100 AFM system. C-V measurement was done with an Agilent E4980A LCR meter. The device electric characterizations were carried out by Keithley 4200 Semiconductor Parameter Analyzer. A Microsanj TR system with high-speed pulsed light emitting diode (LED) ($\lambda$= 530 nm) probe illumination and a synchronized CCD camera was used for the TR measurement equipment.

**Supporting Information**

Details about TR measurement set-up and timing principles, TR coefficient calibration, photography of the TR measurement equipment, simulation mesh build up, heat pulse generation function for transient measurement, simplified 1D equivalent thermal circuit model, and lists of major materials parameters used for thermal simulation and calculation can be found in supporting information.

- **Acknowledgements**

The authors would like to thank Dr. SangHoon Shin for the thermal simulation support and technical guidance from the Sensors Directorate of Air Force Research Laboratory. The work is supported in part by AFOSR (under Grant FA9550-12-1-0180), ONR NEPTUNE (under Grant N00014-15-1-2833), and DTRA (under Grant HDTRA1-12-1-0025).

**Author Information**

Hong Zhou: zhou313@purdue.edu, Kerry Maize: kmaize@purdue.edu, Jinhyun Noh: noh12@purdue.edu, Ali Shakouri: shakouri@purdue.edu, Peide D. Ye: yep@purdue.edu.

**Author contribution**

P. D. Y. conceived the idea and supervised the experiments. H. Z. performed the device fabrication, material and electrical characterization. H. Z. and J. N. performed the thermal simulation. A. S. supervised H. Z. and K. M. to perform the TR characterization. H. Z. and P. D. Y. summarized the manuscript and all authors commented on it.

**Competing financial interests statement**

The authors declare no competing financial interests.

**Figures**

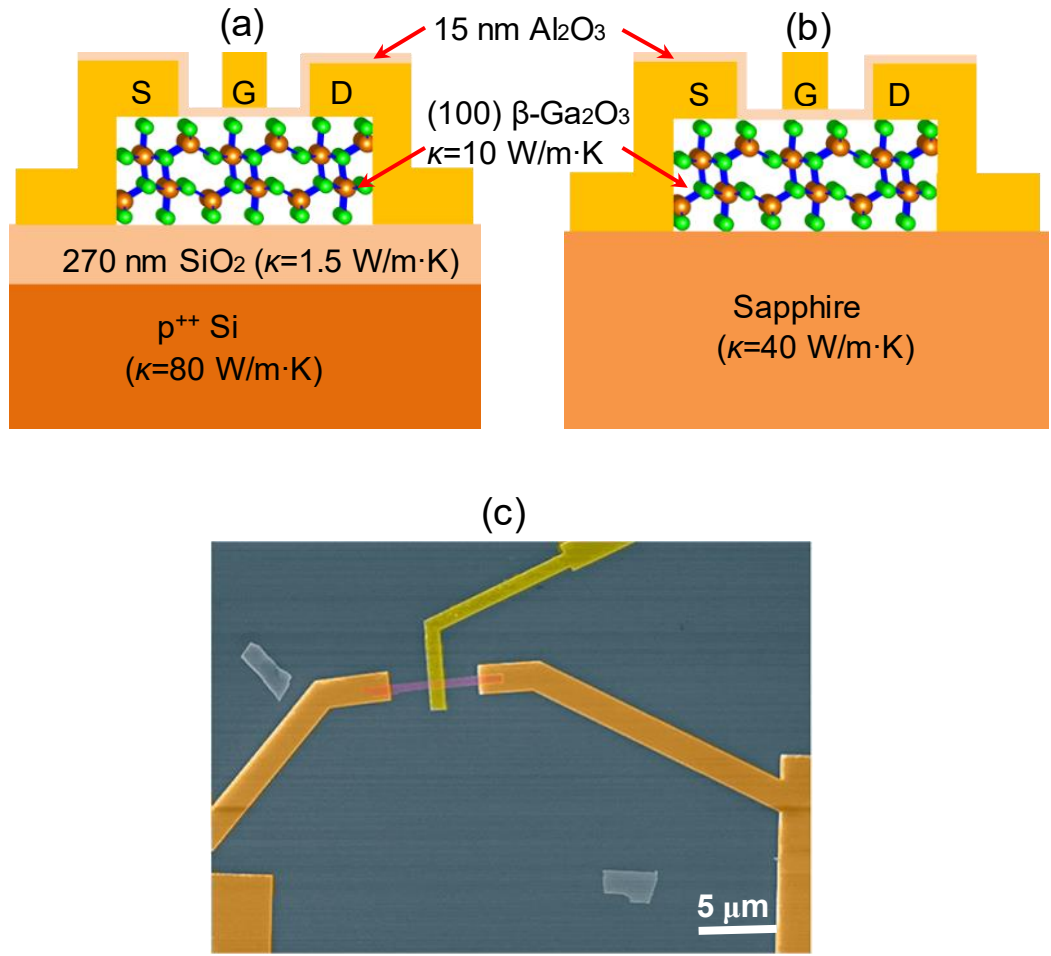

**Figure 1.** Schematic and fabrication of top-gate GOOI FETs. Cross-section schematic view of a top-gate GOOI FET on (a) SiO$_2$/Si and (b) sapphire substrates with different thermal conductivity ($\kappa$) marked. 15 nm of Al$_2$O$_3$ is used as the gate dielectric, Ti/Al/Au (15/60/50 nm) is used as the source and drain electrode, and Ni/Au (30/50 nm) is used the gate electrode. (c) False-colored SEM top-view of a GOOI FET with $L_G$=1 μm and $L_{SD}$=6 μm.

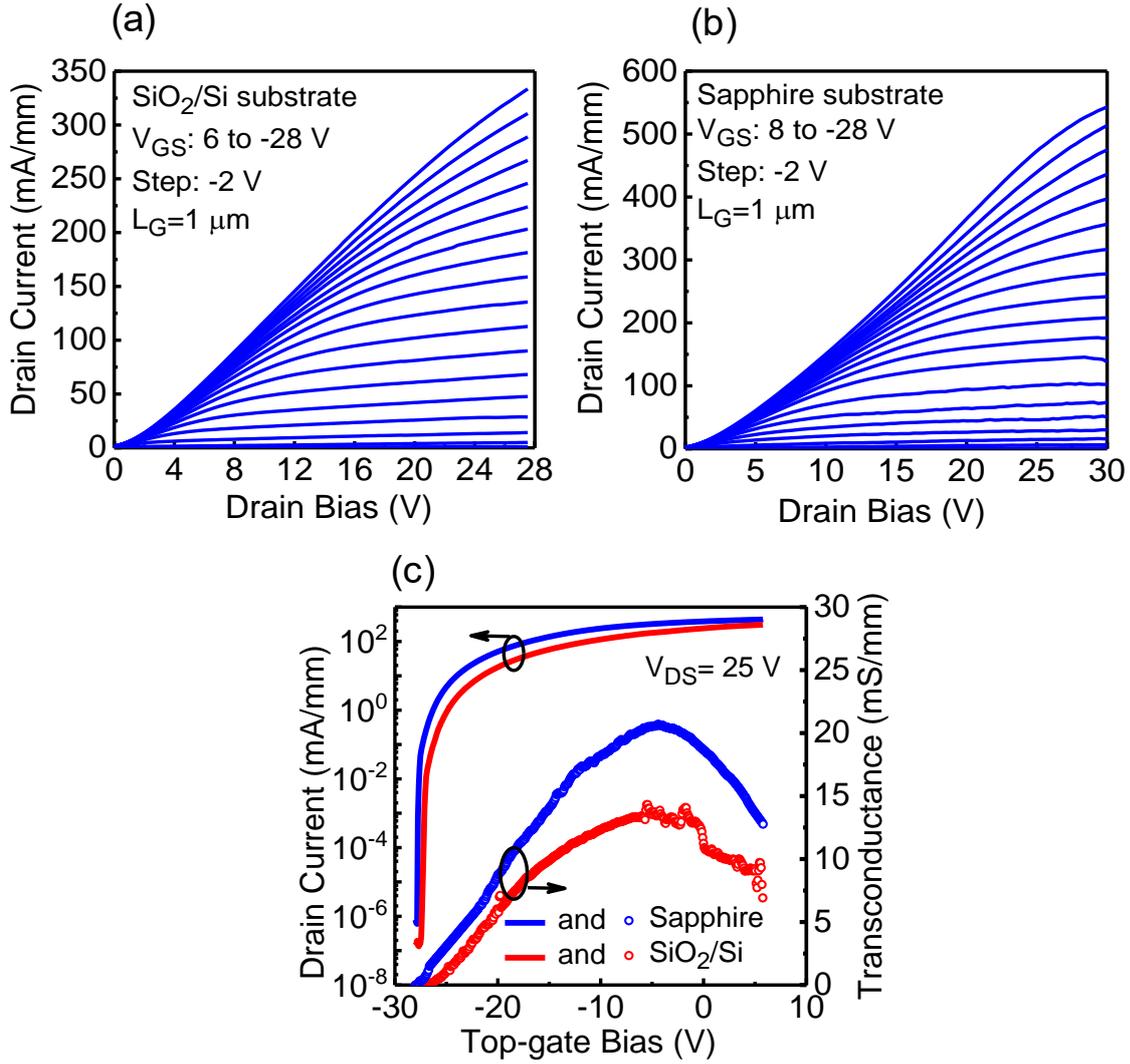

**Figure 2.** I-V electrical characteristics of top-gate GOOI FETs. $I_D$-$V_{DS}$ output characteristics of GOOI FETs on (a) SiO$_2$/Si substrate and (b) sapphire substrate with $L_{SD}$=6~6.5 μm and $L_G$=1 μm. A record high $I_{DMAX}$=535 mA/mm is demonstrated on top-gate β-Ga$_2$O$_3$ GOOI FETs. (c) $I_D$-$g_m$-$V_{DS}$ transfer characteristics of GOOI FETs on both substrates with high on/off ratio of $10^9$ and low SS of 65 mV/dec, yielding a low $D_{it}$ of $2.6\times10^{11}$ eV$^{-1}$·cm$^{-2}$. Higher $g_m$ shows higher electron μ on sapphire substrate due to the reduced device temperature.

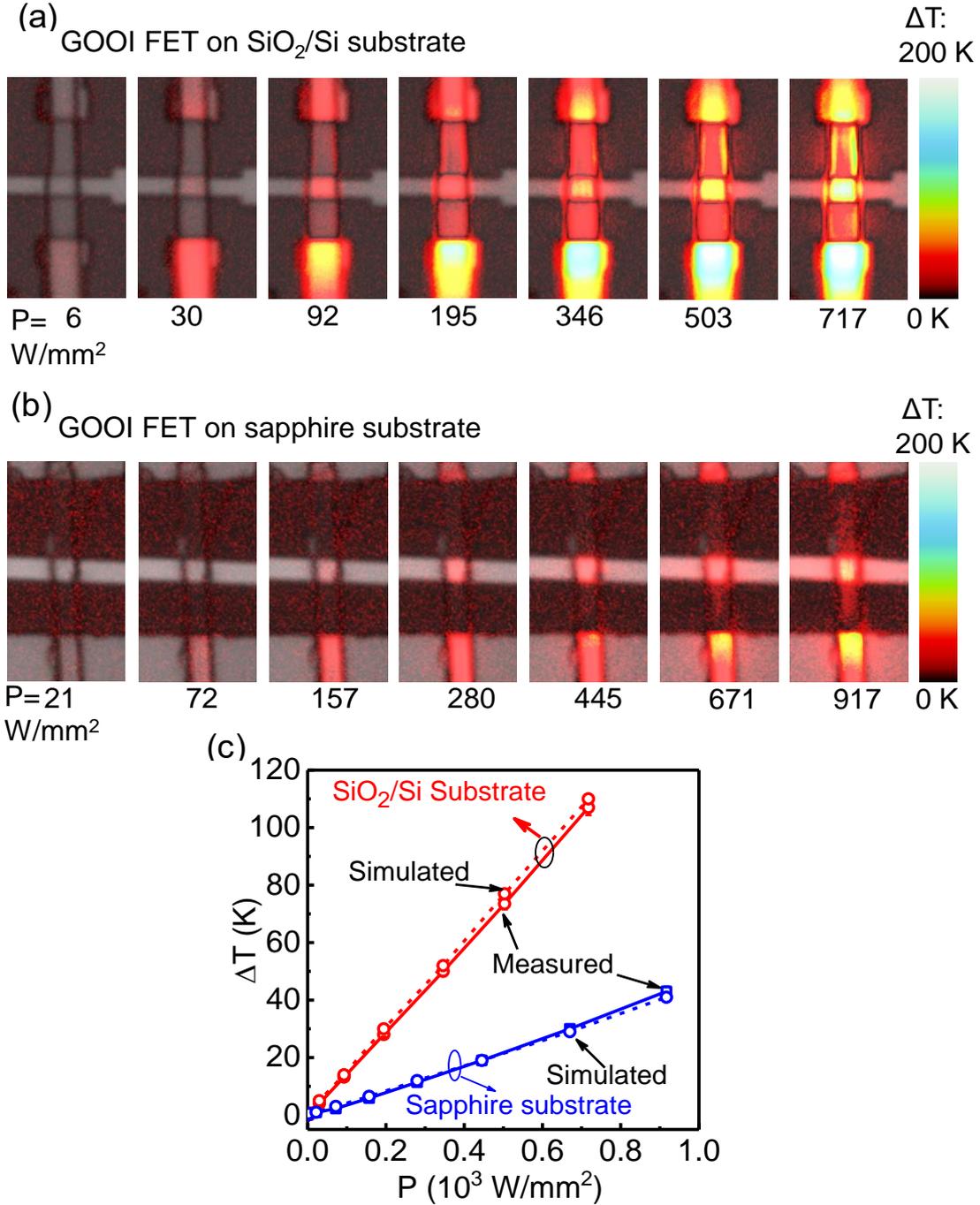

**Figure 3.** Steady state TR characteristics and temperature rise of top-gate GOOI FETs. TR and CCD camera merged images of GOOI FET on (a) $SiO_2$/Si and (b) sapphire substrates when device P is increased by increasing $V_{DS}$ at $V_{GS}$=0 V. Results show device heating at steady state and significant SHE is observed on GOOI FET on $SiO_2$/Si substrate at high P. (c) Measured and simulated local gate Au surface $\Delta T$ comparison between GOOI FET on sapphire and $SiO_2$/Si substrates. Each error bar of measured data is from the standard deviation of 3 devices on both substrates. The power density P=$V_{DS}\times I_D$ is normalized by the β-$Ga_2O_3$ nano-membrane area to avoid random shape induced variation. $\Delta T$ of GOOI FET on $SiO_2$/Si substrate is more than 3 times of that on sapphire substrate at the same P. As a result, the $R_T$ of GOOI FET on sapphire substrate is $4.62\times10^{-2}$ $mm^2$K/W, which is less than 1/3 of that on $SiO_2$/Si substrate.

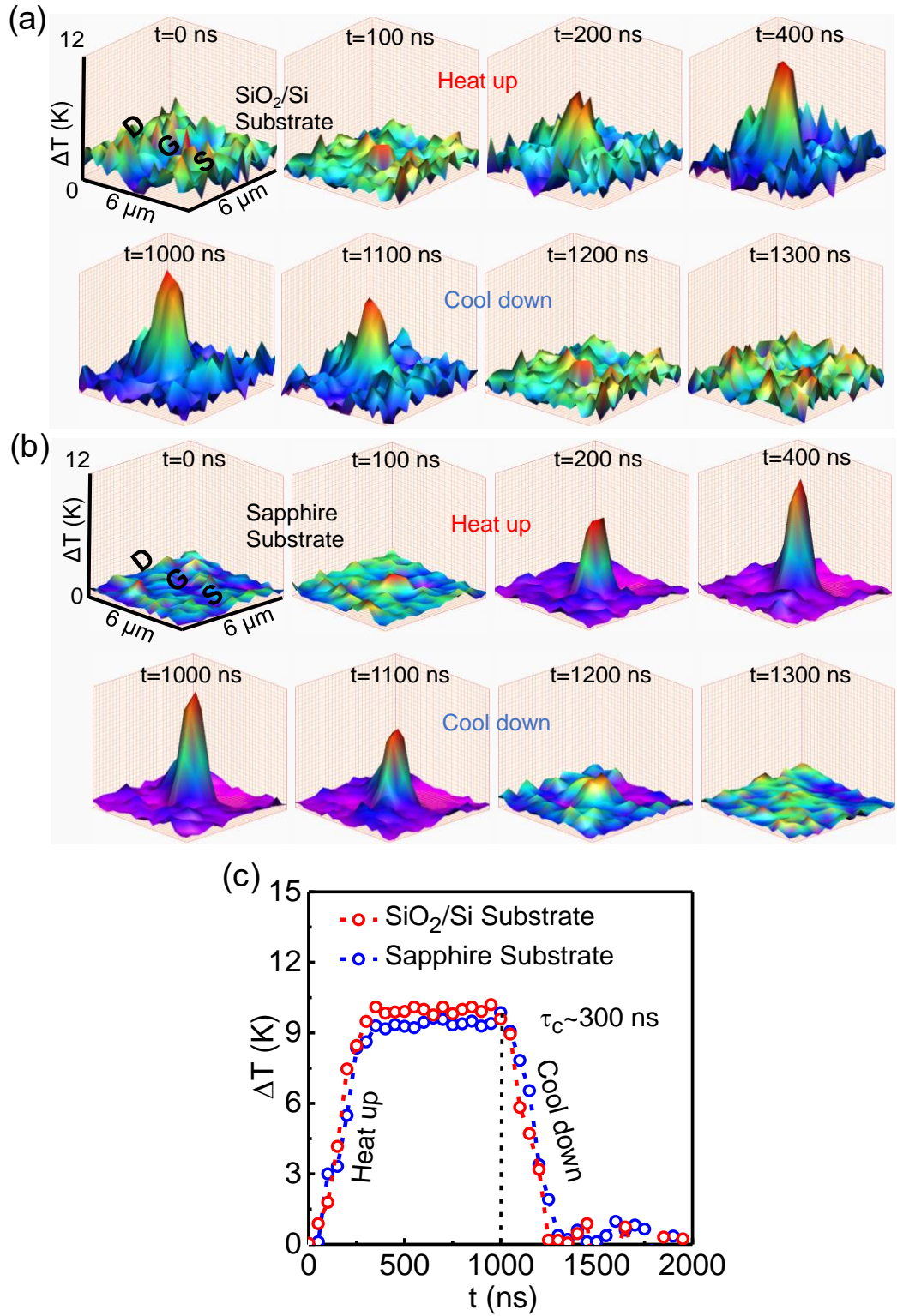

**Figure 4.** Transient TR characteristics of top-gate GOOI FETs. (a) and (b) are 3D TR images of heat up and cool down transient phases of GOOI FETs at gate region on $SiO_2/Si$ and sapphire substrates, respectively. (c) Gate region $\Delta T$ transient comparison between two devices. GOOI FET on sapphire has a cooling phase $\tau_c$ of 300 ns, yielding an overall transient $R_T$ of $5.0\times10^{-2}$ $mm^2K/W$, which is in good agreement with steady state TR measurement.

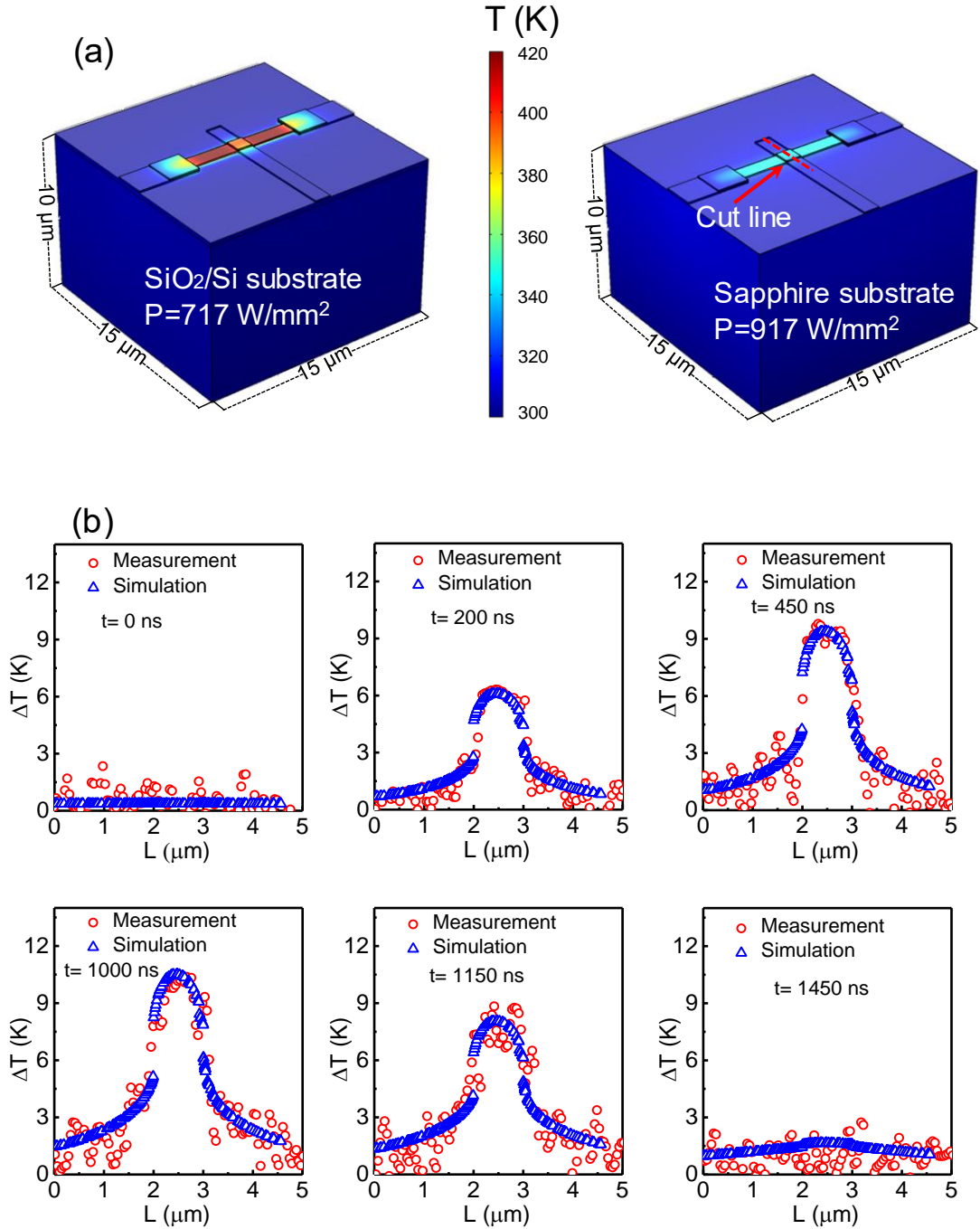

**Figure 5.** Simulated temperature distribution in GOOI FETs and experimental design verification. (a) Simulated temperature distribution of GOOI FETs on $SiO_2$/Si and sapphire substrates with device power density of 717 and 917 $W/mm^2$, respectively. GOOI FET on $SiO_2$/Si has more than twice higher $\Delta T$ than that of sapphire substrate, showing the much severe SHE on $SiO_2$/Si substrate. (b) Simulated and measured temperature distribution along the gate width cutline direction on sapphire substrate at various heat up and cool down phases. Good agreements are achieved both at steady state and transient time domain between measured and simulated data, verifying our experiment design and characterization.

# Table of Content Graphic

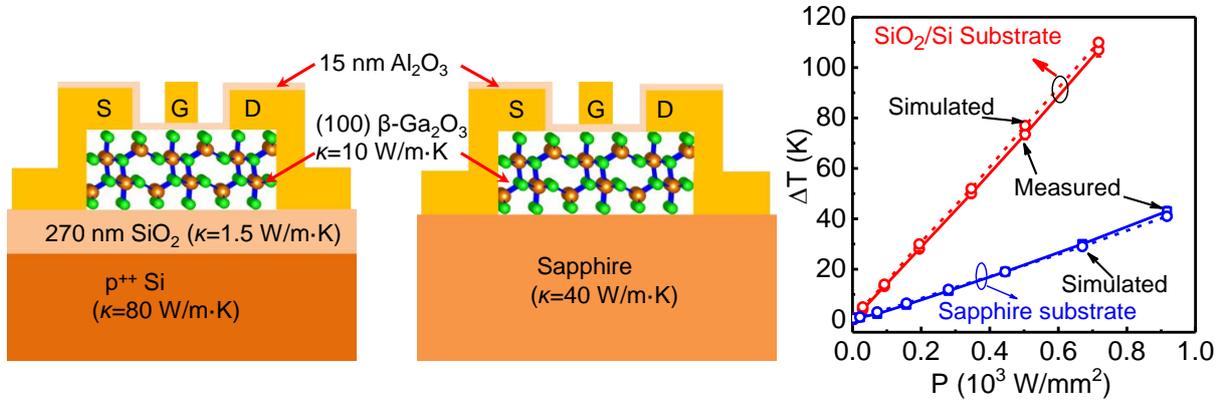